\documentstyle[12pt]{article}
\newcommand{\seq}{Schr\"odinger equation }
\newcommand{\nn}{\nonumber}
\newcommand{\be}{\begin{equation}}
\newcommand{\ee}{\end{equation}}
\newcommand{\bea}{\begin{eqnarray}}
\newcommand{\eea}{\end{eqnarray}}
\newcommand{\r}{\rho}

\newcommand{\lb}{\label}
\newcommand{\rf}{\ref}
\def\lb#1{\label{eq:#1}}
\def\rf#1{(\ref{eq:#1})}

\begin{document}

\baselineskip=14pt plus 0.2pt minus 0.2pt
\lineskip=14pt plus 0.2pt minus 0.2pt

\begin{flushright}
 hepth@xxx/9405154 \\
 LA-UR-94-1374 \\
\end{flushright}

\begin{center}
\Large{\bf  QUANTUM BOUND STATES WITH ZERO BINDING ENERGY}
\vspace{0.25in}

\large

\bigskip

Jamil Daboul\footnote{Email:  daboul@bguvms.bgu.ac.il}\\
{\it Physics Department, Ben Gurion University of the Negev\\
Beer Sheva, Israel}\\
$~~~~~~~$\\
and\\
$~~~~~~~$ \\
Michael Martin Nieto\footnote{Email:  mmn@pion.lanl.gov}\\
{\it
Theoretical Division, Los Alamos National Laboratory\\
University of California\\
Los Alamos, New Mexico 87545, U.S.A. }

\normalsize

\vspace{0.3in}

{ABSTRACT}

\end{center}


\begin{quotation}

After reviewing the general properties of zero-energy quantum
states, we give the explicit solutions of the \seq with $E=0$ for the
class of potentials $V=-|\gamma|/r^{\nu}$, where $-\infty <
\nu < \infty$.  For $\nu > 2$, these solutions are normalizable and
correspond to bound states, if the angular momentum quantum number $l>0$.
[These states are normalizable, even for $l=0$, if we increase the
space dimension, $D$, beyond 4; i.e. for $D>4$.]
For $\nu <-2$ the above solutions, although unbound, are normalizable.
This is true even though
the corresponding potentials are repulsive for all $r$.
We discuss the physics of these unusual effects.


\end{quotation}


\newpage

\section{Introduction}

In studying quantum, nonconfining, potential systems, care is given to
describing both the discrete, normalizable, bound states, which exist
for  energy  $E<0$, and also the non-normalizable, free (including
resonant) states, with energy $E>0$.  However,
usually little is said about
the zero energy states.  From systems  such as the Coulomb problem,
where the $E=0$ state is in the continuum, it can easily be
assumed that all $E=0$ states are  in the continuum and   not normalizable.
However, there are at least two known examples where, for discrete
values of the coupling constant, the  $E=0$ state is bound.

It is the purpose of this note to explore this phenomenon.  We will
demonstrate an exactly solvable system of power-law potentials
where the   $E=0$ states are  bound for {\it continuous}
values of the coupling constant.
We will  illucidate the physics of this situation and also demonstrate that
there also exist normalizable states which cannot be interpreted as bound
states.  Finally,
we will show that by increasing the dimensionality of the problem,
an effective centrifugal barrier is created which causes states to be
bound, even if the expectation value of the angular momentum operator,
$L^2$, vanishes.

\section{Background}

Consider the radial Schr\"odinger equation with angular-momentum
quantum number $l$:
\be
E R_l =\left[-\frac{\hbar^2}{2m}\left(\frac{d^2}{dr^2}
+ \frac{2}{r}\frac{d}{dr}-\frac{l(l+1)}{r^2}\right)
+ V(r)\right] R_l ~.  \label{scheq}
\ee
For the Coulomb problem, the effective potential
\be
U(r) = \frac{\hbar^2}{2m}\frac{l(l+1)}{r^2}
            -\frac{e^2}{r}
\ee
has the form shown in Figure 1.  The effective potential asymptotes to
zero  from below as  $r \rightarrow \infty$, so that a particle with zero
binding energy, $E=0$, has a positive kinetic energy and  is
free to travel  out to infinity.  [Note that in
this case the zero-energy solution, in addition to being continuously connected
to the continuum, is also a limit point of the bound states, whose energies,
$E_n$, go as $-1/n^2$.]

The physical situation is very different, however, if the potential
approaches zero  from the top as $r \rightarrow \infty$.  This is the
case, for example, in the ``standard" discussion of alpha decay.
Consider a phenomenological description of
alpha-decay with a Morse potential.  Then, the
effective potential, with the angular-momentum barrier included,  is
\be
U={\cal E}_0 \left[\frac{l(l+1)}{\r^2} + D\left(-2e^{[-2b(\r-\r_0)]}
                + e^{[- b(\r-\r_0)]}\right)\right]~,
\label{morse}
\ee
where here, and later, we  use the notation,
\be
\r \equiv \frac{r}{a}~,~~~~~{\cal E}_0 \equiv \frac{\hbar^2}{2ma^2}~,
\ee
$a$ being a distance scale and ${\cal E}_0$ being an energy scale.
In Figure 2 we plot an example of the Morse potential.

Early quantum-mechanical text books \cite{mess} discussed
the  energy ranges $E>0$ and $E<0$  for
this type of potential, but often did not include the $E=0$ case
in the discussion.  Even in the famous lecture notes of Fermi
\cite{fermi} the WKB tunnel time to the outside was discussed
only for $E>0$, even though one can see that it goes to infinity as
$E \rightarrow 0$.  [This last argument provides an
intuitive understanding of the bound-state result we are discussing.]

Now we make this point analytic by considering  a solvable system whose
effective potential has the features of the alpha-decay potential.
This potential is
\be
U(r) = \left\{ \begin{array}{lc}
              \frac{\hbar^2}{2m}
             \frac{l(l+1)}{r^2}  - V_0=
{\cal E}_0\left[\frac{l(l+1)}{\r^2}- g^2\right]~,
                &  \mbox{ $r < a$}~, \\
                 ~~~ & ~~~ \\
               \frac{\hbar^2}{2m}
                 \frac{l(l+1)}{r^2}={\cal E}_0\frac{l(l+1)}{\r^2}~,
    &  \mbox{ $r > a$}~.
         \end{array}   \right.   \lb{swell}
\ee
where $g$ is a dimensionless ``coupling constant." In this case, the
effective potential, shown in Figure 3, is
infinite at $r=0$, falls below zero, rises above zero at
some $r=a$, and
then goes to zero from above as $r \rightarrow \infty$.
This is the spherical box, discussed in many places.

The $E=0$ solution is even simpler than the general
case.  To our knowledge, this was the first example of a bound
state with zero binding energy being explicitly demonstrated for a
wide audience \cite{long,longsch,schiff2}.

First consider the interior, $r<a$. The solution is a spherical Bessel
function, which insures that the wave function is finite at the origin:
\be
R_l(r<a) \sim j_l(\kappa r)=j_l(g\r)~, ~~~~\kappa =
\sqrt{2mV_0/\hbar^2}=g/a~, \lb{inside}
\ee
which is $\sim r^{-1/2} J_{(l+1/2)}$.  From the Schr\"odinger equation,
the exterior solutions  $(r>a$, where
$V = 0$), go as a power law.  The choice is the negative power law
since we are interested in normalizable solutions:
\be
R_l(r>a) \sim 1/r^{l+1} ~. \lb{outside}
\ee

The matching condition at $r=a$ is that
\be
\frac{d\ln(R_l)}{dr} = \frac{1}{R_l}\frac{dR_l}{dr}
\ee
be continuous at $r=a$.  This means that
\be
0 = (l+1)j_l(\kappa a) +(\kappa a) j^{\prime}_l(\kappa a)
     = (\kappa a)j_{l-1}(\kappa a)~.
\ee
The first equality is the physical condition.  The second equality is a
standard mathematical result of spherical Bessel functions.

Therefore, the spherical well is a different situation than
the Coulomb case,
where  $E=0$
{\it is} a limit point of the bound-state spectrum.   The spherical
well has a finite bound-state spectrum.  In general
the $E=0$ solutions of the spherical well
are not normalizable.  However, for a given $l$,
an $E=0$ solution is normalizable if $V_0$ is such that
$\kappa a$ is equal to a zero of the spherical Bessel function $j_{l-1}$.
That is,
\be
j_{l-1}(\kappa a) = 0 ~.
\ee
These zeros can easily be calculated and are in tables \cite{sphbes}.

Indeed, in Fig. 3 we plot the effective potential $U(r)$ in \rf{swell}
in units of ${\cal E}_0$ for
$l=1$ and  $V_0=\pi^2 {\cal E}_0$
or $g=\pi$.  Then,  an $E=0$ bound state
exists and corresponds to the first zero, $\kappa a=\pi$, of $j_0(\kappa a)=
\sin (\kappa a)/(\kappa a) $.

Another example of this type is the focusing potential of
Demkov and Ostrovskii \cite{DO,barut}, here written in the form
\be
V = -\frac{w{\cal E}_0 }{\r^2[\r^{\kappa}+\r^{-\kappa}]^2}~, ~~~~ \kappa>0~,
\ee
where $w$ is a
dimensionless coupling constant. This system has normalizable,
 $E=0$  solutions only for the following
discrete values of $w$ \cite{DO,barut}:
\be
w_N=4\kappa^2\left(N +\frac{1}{2\kappa}-1\right)
    \left(N +\frac{1}{2\kappa}\right)~,
\ee
\be
N= n+\left(\frac{1}{\kappa}-1\right)l,~~~~n = n_r+l +1,~~~~
n_r=0,~1,~2,\dots
\ee


\section{Zero-Energy Bound States and Singular Discrete States
for Power-Law Potentials}

Now we present an infinite
class of potentials which is exactly solvable for $E=0$,
 and has  the property that many of the $E=0$ states
are bound.  Elsewhere we will go into more detail on this system
for both the classical case \cite{dn1} and the quantum
case \cite{dn2}.

For convenience we parametrize these potentials as
\be
V(r) = ~-~\frac{\gamma}{r^{\nu}}=
 ~-~\frac{ g^2 {\cal E}_0  }{\r^{\nu}}~,
         ~~~~~~-\infty <\nu < \infty~, \lb{ourpot}
\ee
where  $g$ is a dimensionless coupling constant.
It will be useful to interchange the variables $\nu$ and
$\mu$, which are related by
 \be
\frac{\nu -2}{2} = \mu, ~~~~~\nu = 2(\mu +1)~.
\ee
In Figure 4 we show an example of such a potential where the $E=0$
solution will be a bound state ($\nu > 2$).

We now demonstrate that the Schr\"odinger equation is exactly solvable
for all  $E=0$ and all $-\infty <\nu < \infty$.
 To do this, set $E =0$ in Eq. (\ref{scheq}),
change variables to $\r$,
and then multiply by $-\r^2$.  One finds
\be
0=\left[\r^2\frac{d^2}{d\r^2} + 2\r\frac{d}{d\r} - l(l+1)
         + \frac{g^2}{\r^{2\mu}}\right]R_l(r) ~.
\label{qmsp}
\ee

The above is a  well-known differential equation of
mathematical physics \cite{mos}.  For
$\nu \neq 2$ or $\mu \neq 0$, the solution can be directly
given as
\be
R_l(r) = \frac{1}{\r^{1/2}}
           {\mbox{\huge{J}}}
           _{\left(\frac{2l+1}{|\nu - 2|}\right)}
           \left(\frac{2g}
           {|\nu - 2|\r^{\left(\frac{\nu - 2}{2}\right)}}
           \right)
     = \frac{1}{\r^{1/2}}
           {\mbox{\huge{J}}}
           _{\left(\frac{l+1/2}{|\mu|}\right)}
           \left(\frac{g}
           {|\mu| \r^{\mu}}\right)~, ~~~ \mu \neq 0~.  \label{gensol}
\ee
The other possible solution of
Eq. (\ref{qmsp}), involving the
functions $Y$,  is ruled out
on physical grounds.  (See the Appendix of Ref. \cite{dn2}.)
Also, note that the power in the argument of the solution is {\it not}
an absolute value of $\mu$.  (The singular, free, $\nu = 2$ or $\mu = 0$
case will be discussed in Ref. \cite{dn2}.)

We now find out under what circumstances
these states are normalizable. The normalization constants for the wave
functions would have to be of the form
\be
N_l^{-2} = \int_{0}^{\infty}\frac{r^2dr}{\r}
             {\mbox{\huge{J}}}^2
           _{\left(\frac{l+1/2}{|\mu|}\right)}
           \left(\frac{g}
           {|\mu| \r^{\mu}}\right)~.\label{norm}
\ee
Changing variables first from $r$ to $\r$, and then from $\r$ to
$z=g/(|\mu|\r^{\mu})$, and being
careful about the limits of integration for all $\mu$, one obtains
\be
N_l^{-2} = \frac{a^3}{|\mu|}\left(\frac{g}{|\mu|}\right)^{2/\mu}~
I_l~~,
\ee
where
\be
I_l = \int_0^{\infty}\frac{dz}{z^{(1+2/\mu)}}
              {\mbox{\huge{J}}}^2
           _{\left(\frac{l+1/2}{|\mu|}\right)}
           (z) ~. \lb{int}
\ee

Integrals of products of Bessel functions are well studied and
are  complicated
when the orders and arguments approach each other \cite{wat}.
However, this   integral
 is convergent and given by \cite{wat2}
\be
I_l= \frac{1}{2\pi^{1/2}}
\frac{\Gamma\left(\frac{1}{2}+\frac{1}{\mu}\right)}
{\Gamma\left(1+\frac{1}{\mu}\right)}
\frac{\Gamma\left(\frac{l+1/2}{|\mu|}-\frac{1}{\mu}\right)}
{\Gamma\left(1+\frac{l+1/2}{|\mu|}+\frac{1}{\mu}\right)}~,
\lb{norm}
\ee
if the following  two conditions are satisfied:
\be
\frac{2l+1}{|\mu|} + 1> \frac{2}{\mu}+ 1>0~.  \lb{gt}
\ee
(In obtaining the final result in Eq. \rf{norm}
the doubling formula for $\Gamma(2z)$  was used.)

Eqs. \rf{norm} and \rf{gt} lead to two sets of
normalizable states.   The first is when
\be
\mu > 0 ~~~ \mbox{or} ~~~\nu > 2~,   ~~~~~~~  l >1/2~.
\lb{lll}
\ee
  These are ordinary
 bound states and result because the effective
 potential asymptotes to zero from above, as in Figure 4.
In this case, for $E=0$, the wave function can reach infinity
only by tunneling through an infinite forbidden region.
 That takes forever, and so the state is bound.
Note that the condition on $l$
in Eq. \rf{lll} is the minimum nonzero angular momentum
allowed in quantum mechanics, $l_{min} = 1$.  This agrees with the
classical orbit solution which is bound for any nonzero
angular momentum \cite{dn1}.

 Notice that the above $E=0$ solutions exist for all $g^2 > 0$, and not
just for discrete values of the coupling constant. The reason for this
surprising result is the
scaling property of  power-law potentials.  A change of the coupling
constant, $g^2$, by a positive factor, to $\sigma^2 g^2$,  can be
accounted for by changing the argument of the wave function from
$g/|\mu|\r^\mu$ to $\sigma g/|\mu|\r^\mu$.  This is essentially a change of
the length scale.

For $-2 \leq \nu \leq 2$ or $-2 \leq \mu \leq 0$ (as well as the
solutions with $l=0$ and $\mu > 0$ or $\nu > 2$) the solutions are
free, continuum solutions.

However, there is one remaining class of
normalizable solutions which is quite surprising.
For any $l$ and all $\nu < -2$ or $\mu<-2$, the reader can
verify that  the conditions of Eq. \rf{gt} are also satisfied.
Thus, even though one here has a repulsive potential that falls
off faster than the inverse-harmonic oscillator and the
states are {\it not} bound, the solutions are
normalizable!

The corresponding classical solutions yield
infinite orbits, for which the particle needs only a finite time
to reach infinity \cite{dn1}.
But it is known that a classical potential which yields
trajectories with a finite travel time to infinity
also yields a discrete spectrum
in the quantum case \cite{wightman}.
This conclusion is in agreement with the
situation here.  Although   normalizable quantum solutions exist
not just for $E=0$ but also for a continuous range of $E$,
by imposing special boundary conditions a discrete subset can
be chosen which defines a self-adjoint extension of the
Hamiltonian \cite{case,berezin,klauder}.

This system has many other interesting features, both
classically and quantum mechanically.  We refer the reader elsewhere
to discussions of these properties \cite{dn1,dn2}.


\section{Bound States in Arbitrary Dimensions}

One can easily generalize the problem of the last
section to arbitrary $D$
space  dimensions.   Doing so yields another surprising
physical result.

To obtain the
$D$-dimensional analogue of Eq. (\ref{qmsp}),  one simply has to
replace $2\r$ by $(D-1)\r$ and $l(l+1)$ by $l(l+D-2)$ \cite{nd}:
\be
0=\left[\r^2\frac{d^2}{d\r^2} + (D-1)\r\frac{d}{d\r} - l(l+D-2)
        +\frac{g^2}{\r^{2\mu}}\right]R_{l,D}~.
\lb{qmD}
\ee
The solutions also follow similarly as
\bea
R_{l,D} &=& \frac{1}{\r^{D/2-1}}
           {\mbox{\huge{J}}}
           _{\left(\frac{2l+D-2}{|\nu - 2|}\right)}
           \left(\frac{2g}
           {|\nu - 2|\r^{\left(\frac{\nu - 2}{2}\right)}}
           \right)
                        \\ \nonumber
     &=& \frac{1}{\r^{D/2-1}}
           {\mbox{\huge{J}}}
           _{\left(\frac{l+D/2-1}{|\mu|}\right)}
           \left(\frac{g}
           {|\mu| \r^{\mu}}\right)~.
\eea

To find out which states are normalizable one first has to change the
integration measure  from $r^2dr$ to $r^{D-1}dr$ and again
proceed as before. The end
result is that if the wave functions are normalizable,
the normalization constant is given by
\be
N_{l,D}^{-2} =\frac{a^D}{|\mu|}\left(\frac{g}{|\mu|}\right)^{2/\mu}~
I_{l,D}~~,
\ee
where
\be
I_{l,D} =  \int_0^{\infty}\frac{dz}{z^{(1+2/\mu)}}
              {\mbox{\huge{J}}}^2
           _{\left(\frac{l+D/2-1}{|\mu|}\right)}
           (z) ~. \lb{ild}
\ee
We see that the above integral is equal exactly to that in \rf{int},
except that $l$ is replaced by the effective quantum number
\be
l_{eff}=l+\frac{D-3}{2}~. \lb{leff}
\ee
Therefore,
\be
I_{l,D} = \frac{1}{2\pi^{1/2}}
\frac{\Gamma\left(\frac{1}{2}+\frac{1}{\mu}\right)}
{\Gamma\left(1+\frac{1}{\mu}\right)}
\frac{\Gamma\left(\frac{l+D/2-1}{|\mu|}-\frac{1}{\mu}\right)}
{\Gamma\left(1+\frac{l+D/2-1}{|\mu|}+\frac{1}{\mu}\right)}~,
\ee
which is defined and convergent for
\be
\frac{2l+D-2}{|\mu|}+1 >\frac{2}{\mu}+1>0~.
 \lb{gtD}
\ee

This yields the surprising result that there are bound states for
all $\nu > 2$ or $\mu > 0$ when $l >2-D/2$.  Explicitly this
means that the minimum allowed $l$ for there to be zero-energy
bound states are:
\bea
D = 2~, ~~~~~ l_{min} &=& 2~, \nonumber \\
D = 3~, ~~~~~ l_{min} &=& 1~, \nonumber \\
D = 4~, ~~~~~ l_{min} &=& 1~, \nonumber \\
D > 4~, ~~~~~ l_{min} &=& 0~. \lb{lmin}
\eea

This effect of dimensions is purely quantum mechanical and can be
understood as follows:  Classically, the number of dimensions involved
in a central potential problem has no intrinsic effect on the
dynamics.  The orbit remains in two dimensions, and the problem is
decided by the form of the effective potential, U, which contains
only the angular momentum barrier and the dynamical potential.

In quantum mechanics there are actually two places where
an effect of dimension appears.  The first is in the factor $l(l+D-2)$
of the angular-momentum
barrier.  The second is more
fundamental. It is due to the operator
\be
U_{qm} = -\frac{(D-1)}{\r}\frac{d}{d\r}~.
\ee
The contribution of $U_{qm}$ to the ``effective potential"
can be calculated by using the ansatz
\be R_{l,D}(\r)\equiv \frac{1}{\r^{(D-1)/2}}\; \chi_{l,D}(\r)~. \lb{ansatz}
\ee
This transforms the $D-$dimensional radial \seq into a
$1-$dimensional Schr\"odinger equation:
\be
0=\left[-\frac{d^2}{d\r^2}+ U_{l,D}(\r)\right] \chi_{l,D}~.
\lb{qmDeff}
\ee
In Eq. \rf{qmDeff}, the effective potential $U_{l,D}(\r)$ is given by
\bea
U_{l,D}(\r)&=&\frac{(D-1)(D-3)}{4 \r^2} + \frac{l(l+D-2)}{\r^2}+
V(\r)\nn \\
&=&
\frac{l_{eff}(l_{eff}+1)}{\r^2}+V(\r)~,
\lb{udeff}
\eea
with $l_{eff}$ given in Eq. \rf{leff}. Since the \seq \rf{qmDeff} depends
only on the combination
$l_{eff}$, the solution $\chi_{l,D}(\r)$ does not depend
on $l$ and $D$ separately.
This explains, in particular, the values of $l_{min}$ given in
Eq.  \rf{lmin}.

Although the above ansatz is well known, the
dimensional effect has apparently not been adequately appreciated.
One reason may be attributed to the fact that in going from $D=3$ to
$D=1$, $l_{eff}$ remains equal to $l$.
However, in our problem this effect leads to such a
counter-intuitive result, that it cannot be overlooked.

The dimensional effect essentially produces an additional
centrifugal barrier
which can bind the
wave function at the threshold, even though the
expectation value of the angular momentum vanishes.  Note that this
is in distinction to the classical problem, where there would be no
``effective" centrifugal barrier to prevent the particle from
approaching $r \rightarrow \infty$.


\section{Summary}
After obtaining exact, $E=0$ solutions of the \seq  for power-law
potentials, we demonstrated three interesting effects: \\

\noindent (1) There exist
bound states at the threshold, {\em for all $l>0$ and
all $\gamma >0$}. These states  persist if one changes the coupling
constant $\gamma$  by a positive factor.
[In contrast,  $E=0$ bound states exist
for the spherical well and the
focusing potentials
only for very special values of the coupling constants,
 and never for all $l>0$ simultaneously.]   \\

\noindent (2) There exist normalizable solutions for $\nu<-2$, i.e., for
highly repulsive potentials, singular at $\r \rightarrow \infty$. \\

\noindent (3) For higher-space dimensions, each additional
dimension adds a half unit to the effective angular-momentum
quantum number, $l_{eff}$, of Eq. \rf{leff}.
An  effective centripetal barrier,  solely due to this dimensional
effect,  i.e., for $L^2=0$,  is capable of producing a bound state.
This result is a  remarkable manifestation of quantum mechanics
and  has no classical counterpart.


\newpage


{\bf Figure Captions} \\

Figure 1:  A dimensionless representation of the Coulomb effective
potential.  We show the effective potential $U=1/\r^2 - 4/\r~.$  \\

Figure 2:  The Morse effective potential of Eq. (\ref{morse}) in units of
${\cal{E}}_0 = \hbar^2/(2ma^2)$,
We take $l=2$, $D=11$, $b=2.5$, and $\r_0 =1$.  \\

Figure 3.  The effective potential $U(\r)$ of Eq. \rf{swell},
in units of ${\cal E}_0$, for a
spherical-well, with $l=1$ and $V_0 =\pi^2 {\cal E}_0$ or
$g=\pi$. With these parameters,
there is exactly one bound state, at $E=0$.  \\

Figure 4:  The effective potential obtained from Eq. \rf{ourpot}
for $\nu = 4$ in units of ${\cal E}_0/2$, as a function of $\r = r/a$.
The form is $U(\r)=4/\r^2 - 1/\r^4$.  \\

\end{document}